**Технические науки**

# Установка и апробация серверной компоненты информационной образовательной среды университета на платформе LMS Moodle

[1] Ирина Ильинична Ерёмина
[2] Александр Куртович Розенцвайг
[3] Рушан Анурович Зиатдинов

[1-2] Набережночелнинский институт, Казанский федеральный университет,
г. Набережные Челны, Республика Татарстан, Российская Федерация
[1] Кандидат педагогических наук, доцент
E-mail: ereminaii@yandex.ru
[2] Доктор технических наук, профессор
E-mail: a_k_r@mail.ru
[3] Университет Фатих, г. Стамбул, Турция
Кандидат физико-математических наук, профессор-ассистент
E-mail: ziatdinov@fatih.edu.tr, rushanziatdinov@yandex.ru

**Аннотация.** Информационная образовательная среда (ИОС) учебного заведения представляет собой сложную многоуровневую систему и наряду с программно-методическими, организационными и культурными ресурсами аккумулирует интеллектуальный, технический потенциал ВУЗа, содержательные и деятельностные компоненты обучаемых и педагогов. На практике формирование ИОС фактически основывается на создании информационных технологий и интеграции их в существующие образовательные среды учебного заведения. При этом управление такой системой осуществляется специализированным оборудованием и программными средствами. Для успешного формирования и функционирования ИОС в настоящей работе рассматриваются программные продукты, составляющие основу организации интерактивного и веб-взаимодействия студентов, преподавателей и всех участников учебного процесса. Анализируются технические возможности, которые предоставили пользователям ИОС такие сервисы, как веб-сервер Apache с подключенными модулями PHP, СУБД MySQL, виртуальная машина Java и Red5 Server. Демонстрируются возможности получения результатов взаимодействия этих продуктов, отчетов о работе пользователей в вебинарах, видеоконференциях и веб-конференциях.

**Введение.** Новые требования к подготовке будущих специалистов в условиях информатизации в качестве приоритетных направлений модернизации инженерного образования обусловливают поиск новых форм, методов и средств обучения. Интерактивные и сетевые технологии позволяют выносить за пределы учебных аудиторий как теоретическую составляющую изучаемых дисциплин, так и практические занятия и лабораторные работы. В этой связи целесообразно применять комплекс электронных образовательных ресурсов (ЭОР) для организации и осуществления основных этапов самостоятельной информационной учебной деятельности студентов (постановки цели, планирования, осуществления, контроля, коррекции, оценки деятельности).

LMS Moodle – это система управления курсами, также известная как система управления обучением или виртуальная обучающая среда. Она представляет собой свободное (распространяющееся по лицензии GNU GPL) веб-приложение, предоставляющее возможность создавать сайты для онлайн-обучения. Система реализует философию «педагогики социального конструкционизма» и ориентирована, прежде всего, на организацию взаимодействия между преподавателем и учениками, хотя подходит и для организации традиционных дистанционных курсов, а также поддержки очного обучения. LMS Moodle переведена на десятки языков, в том числе и русский и используется почти в 50 тысячах организаций из более чем 200 стран мира. В Российской Федерации зарегистрировано более 1000 инсталляций. Количество пользователей LMS Moodle в некоторых инсталляциях достигает 500 тысяч человек. Лидером и идеологом системы является Martin Dougiamas из Австралии. Проект является открытым и в нём участвует и множество других разработчиков, например, русификацию LMS Moodle осуществляет команда добровольцев из России, Белоруссии и Украины.

**Выбор программного обеспечения и оптимальных параметров технического сопровождения ЭОР**

Для экспериментальной работы по созданию редактированию и внедрению ЭОР в учебный процесс было принято решение установить LMS Moodle сети Набережночелнинского института КФУ, отработать связку «LMS Moodle – Openmeetings» и выбрать оптимальные параметры технического сопровождения. При проектировании ИОС за основу была взята ОС Community ENTerprise Operation System (CentOS). CentOS – операционная система корпоративного использования (один из дистибутивов OS Linux), полученная от поставщика Red Hat Enterprise Linux (RHEL). Наш выбор обусловлен надежностью, направленностью на серверное использование, распространенностью и совместимостью с RHEL. Проанализировав функциональные возможности и область применения веб-сервера Apache и сервера баз данных MySQL, был определен выбор программных продуктов для организации взаимодействия операционной системы CentOS, платформы Moodle и системы организации вебинаров Openmeetings.

В качестве сервера был использован компьютер следующей конфигурации:

- Процессор: Intel Celeron CPU 2.80GHz;
- Оперативная память: 512 MB DDR-II;
- Жесткий диск: Samsung HDD 80GB Sata-II;
- Материнская плата: Intel® D945GTP;
- Сетевая карта: Intel 100Мб/сек.

Установка производилась на отдельный раздел жесткого диска объемом 25 Гбайт. Разбиение было сделано следующим:

- 1,5 Гбайт выделено под файл подкачки (SWAP);
- 100 Мбайт выделено под раздел Boot (/boot);
- 23 Гбайт – под ОС и данные (/).

Установка производилась путем выбора русского языка и русскоязычной раскладки, часового пояса UTC+4 — Московское время, Объединённые Арабские Эмираты, Оман, Азербайджан, Армения, Грузия (Delta).

Для сервера в существующей сети был выделен IP-адрес 172.16.0.28, маска сети 255.255.252.0, имя узла установлено как web-centos.

Для установки были выбраны только пакеты раздела «Базовая система - Основа».

Файервол настроен на запрет входящих подключений к серверу, кроме портов веб-сервера – 80; клиента ssh – 35625 и openmeetings 5080, 1935. Исходящий трафик разрешен. Переадресация отключена.

Сервер ssh настраивался на прием соединений с порта 35625, использование протокола 2 версии, так как 1 версия протокола была успешно взломана. Также был запрещен удаленный вход суперпользователя (root), и разрешен вход только для пользователя adminOS.

В стандартной комплектации дистрибутива входит большое количество служб, для работы веб-сервера многие службы не требуются, поэтому они были отключены. Кроме того, были удалены некоторые компоненты системы, которые не нужны для работы веб-сервера.

Для установки свежих версий пакетов php, веб-сервера Apache, СУБД MySQL и сервера Red5 в систему были добавлены несколько репозиториев. В частности: EPEL (Extra Packages for Enterprise Linux), remi (Les RPM de Remi) и RPMForge.

Установлены пакеты следующих версий: httpd 2.2.3-43, mysql-server 5.1.54-1, php 5.3.5-1, java 1.6.0. Операционная система CentOS версии 5.5, ядро 2.6.18-194.32.1.

В конфигурацию сервера Apache внесены следующие изменения:

- значение ServerTokens установлено Prod, чтобы не показывать в случае ошибки запроса версии ОС, веб-сервера;
- значение TimeOut установлено 45;
- значение KeepAlive установлено «on», разрешает персистентные соединения;
- значение MaxKeepAliveRequest установлено 50, задает максимальное количество запросов при одном персистентном соединении;
- значение KeepAliveTimeout установлено 2, время в секундах ожидания запросов в одном персистентном соединении;
- значение StartServers установлено 4, определяет количество дочерних процессов, запускаемых при запуске сервера Apache;
- значение MinSpareServers установлено 3, задает минимальное количество свободных дочерних процессов;
- значение MaxSpareServers установлено 4, задает максимальное количество свободных дочерних процессов;
- значение ServerLimit установлено 10, устанавливает максимальное значение MaxClients;
- значение MaxCliets установлено 10, задает максимальное число дочерних процессов, которым разрешено запуститься для обработки запросов;
- порт для обслуживания http-запросов установлен в значение 80;
- в значение DirectoryIndex добавлен index.php;
- отключены все языки кроме русского и английского;
- директива ServerName установлена localhost;
- добавлены виртуальные хосты LMS Moodle.ru и testovik.ru.

Значения StartServers, MinSpareServers, MaxSpareServers, MaxClients были установлены исходя из объема оперативной памяти.

В конфигурацию MySQL внесены следующие изменения:

- порт для подключения установлен в значение 3306 (стандартный);
- кодировка по умолчанию была установлена в формат utf8.

Остальные параметры были установлены из файла стандартной конфигурации MySQL, расположенного в директории /usr/share/mysql. По данной ссылке представлены примеры настроек для различного использования сервера MySQL.

Созданы базы данных для LMS Moodle и Openmeetings. Также установлены пользователи этих баз данных и пароли к ним.

Файлы LMS Moodle были размещены в директории /var/www/LMS Moodle.ru, файлы данных находятся в директории /usr/LMS Moodledata.

Сервер Red5 был собран и установлен в директорию /usr/local/red5. Файлы Openmeetings были размещены в директории /usr/local/red5/webapps/openmeetings.

Первоначальное тестирование проводилось в беспроводной сети Wi-Fi. LMS Moodle функционирует стабильно, результаты приведены на Рис. 1.

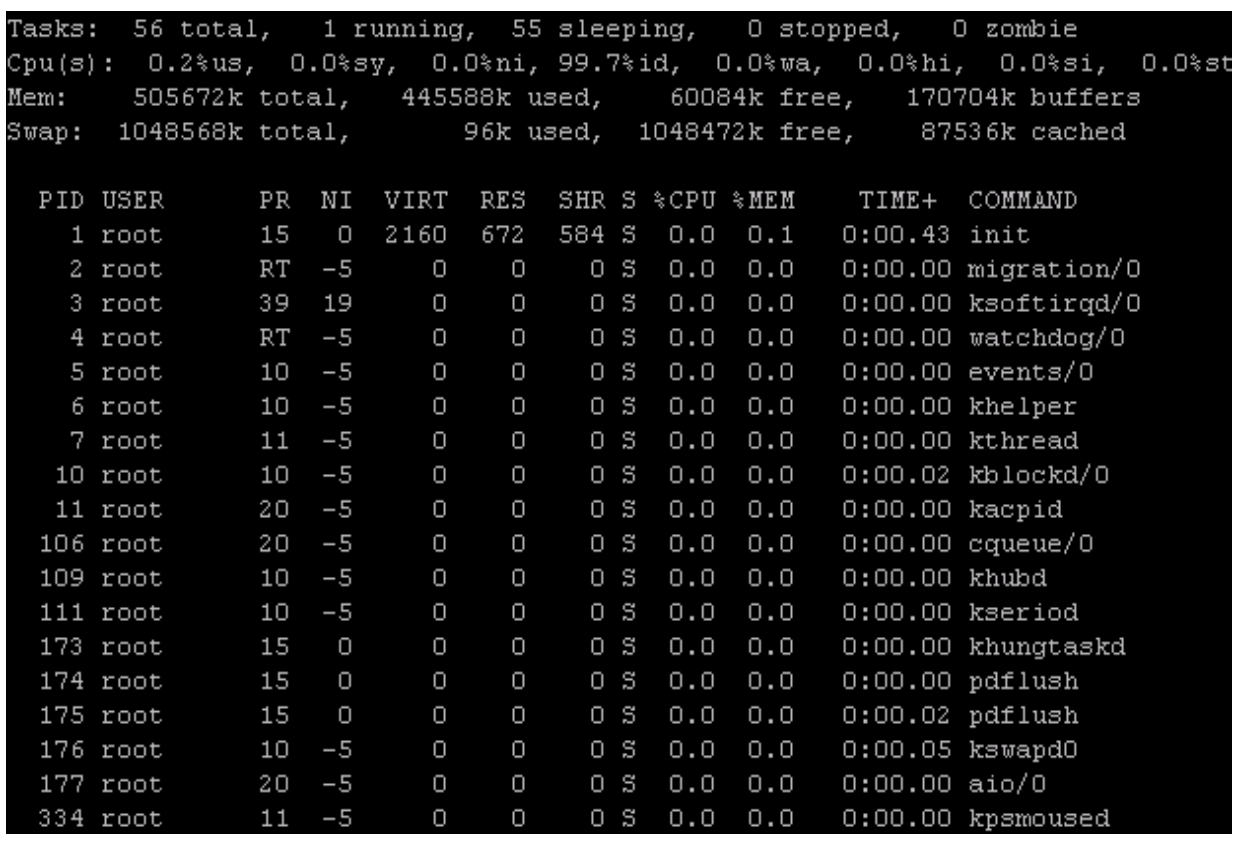

```
Tasks:  56 total,   1 running,  55 sleeping,   0 stopped,   0 zombie
Cpu(s):  0.2%us,  0.0%sy,  0.0%ni, 99.7%id,  0.0%wa,  0.0%hi,  0.0%si,  0.0%st
Mem:    505672k total,   445588k used,    60084k free,   170704k buffers
Swap:  1048568k total,       96k used,  1048472k free,    87536k cached

  PID USER      PR  NI  VIRT  RES  SHR S %CPU %MEM    TIME+  COMMAND
    1 root      15   0  2160  672  584 S  0.0  0.1   0:00.43 init
    2 root      RT  -5     0    0    0 S  0.0  0.0   0:00.00 migration/0
    3 root      39  19     0    0    0 S  0.0  0.0   0:00.00 ksoftirqd/0
    4 root      RT  -5     0    0    0 S  0.0  0.0   0:00.00 watchdog/0
    5 root      10  -5     0    0    0 S  0.0  0.0   0:00.00 events/0
    6 root      10  -5     0    0    0 S  0.0  0.0   0:00.00 khelper
    7 root      11  -5     0    0    0 S  0.0  0.0   0:00.00 kthread
   10 root      10  -5     0    0    0 S  0.0  0.0   0:00.02 kblockd/0
   11 root      20  -5     0    0    0 S  0.0  0.0   0:00.00 kacpid
  106 root      20  -5     0    0    0 S  0.0  0.0   0:00.00 cqueue/0
  109 root      10  -5     0    0    0 S  0.0  0.0   0:00.00 khubd
  111 root      10  -5     0    0    0 S  0.0  0.0   0:00.00 kseriod
  173 root      15   0     0    0    0 S  0.0  0.0   0:00.00 khungtaskd
  174 root      15   0     0    0    0 S  0.0  0.0   0:00.00 pdflush
  175 root      15   0     0    0    0 S  0.0  0.0   0:00.02 pdflush
  176 root      10  -5     0    0    0 S  0.0  0.0   0:00.05 kswapd0
  177 root      20  -5     0    0    0 S  0.0  0.0   0:00.00 aio/0
  334 root      11  -5     0    0    0 S  0.0  0.0   0:00.00 kpsmoused
```

*Рис. 1.* Вид системы в состоянии покоя

В состоянии бездействия параметры системы следующие:

- загруженность процессора менее 1 % от его мощности;
- Из имеющегося объема 512 Мбайт занято 445 Мбайт, свободно 60 Мбайт;
- активен только один процесс init. При загрузке компьютера происходит последовательная передача управления от BIOS к загрузчику, а от него – к ядру. Затем ядро запускает планировщик (для реализации многозадачности) и выполняет программу init. Программа init настраивает пользовательское окружение и позволяет осуществлять взаимодействие с пользователем и вход в систему), после чего ядро переходит в состояние бездействия до тех пор, пока не получит внешний вызов.

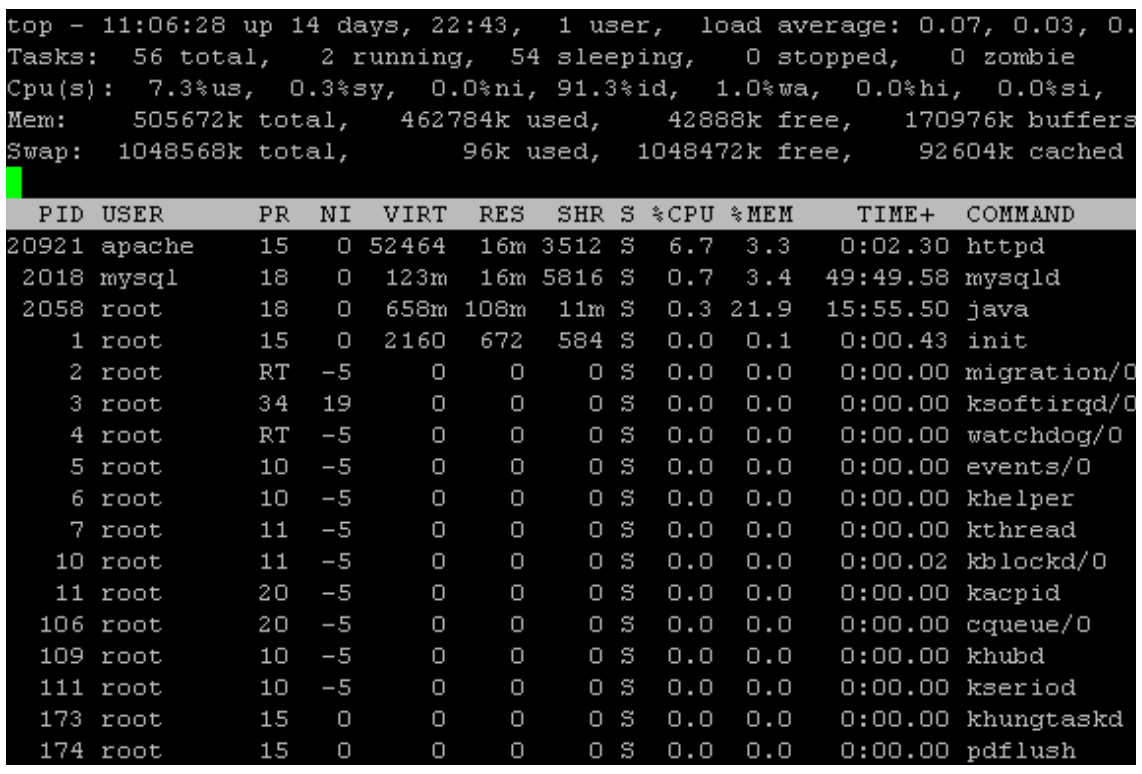

```
top - 11:06:28 up 14 days, 22:43,  1 user,  load average: 0.07, 0.03, 0.
Tasks:  56 total,   2 running,  54 sleeping,   0 stopped,   0 zombie
Cpu(s):  7.3%us,  0.3%sy,  0.0%ni, 91.3%id,  1.0%wa,  0.0%hi,  0.0%si,
Mem:    505672k total,   462784k used,    42888k free,   170976k buffers
Swap:  1048568k total,       96k used,  1048472k free,    92604k cached

  PID USER      PR  NI  VIRT  RES  SHR S %CPU %MEM    TIME+  COMMAND
20921 apache    15   0 52464  16m 3512 S  6.7  3.3   0:02.30 httpd
 2018 mysql     18   0  123m  16m 5816 S  0.7  3.4  49:49.58 mysqld
 2058 root      18   0  658m 108m  11m S  0.3 21.9  15:55.50 java
    1 root      15   0  2160  672  584 S  0.0  0.1   0:00.43 init
    2 root      RT  -5     0    0    0 S  0.0  0.0   0:00.00 migration/0
    3 root      34  19     0    0    0 S  0.0  0.0   0:00.00 ksoftirqd/0
    4 root      RT  -5     0    0    0 S  0.0  0.0   0:00.00 watchdog/0
    5 root      10  -5     0    0    0 S  0.0  0.0   0:00.00 events/0
    6 root      10  -5     0    0    0 S  0.0  0.0   0:00.00 khelper
    7 root      11  -5     0    0    0 S  0.0  0.0   0:00.00 kthread
   10 root      11  -5     0    0    0 S  0.0  0.0   0:00.02 kblockd/0
   11 root      20  -5     0    0    0 S  0.0  0.0   0:00.00 kacpid
  106 root      20  -5     0    0    0 S  0.0  0.0   0:00.00 cqueue/0
  109 root      10  -5     0    0    0 S  0.0  0.0   0:00.00 khubd
  111 root      10  -5     0    0    0 S  0.0  0.0   0:00.00 kseriod
  173 root      15   0     0    0    0 S  0.0  0.0   0:00.00 khungtaskd
  174 root      15   0     0    0    0 S  0.0  0.0   0:00.00 pdflush
```

*Рис. 2.* Вид системы (работает один пользователь)

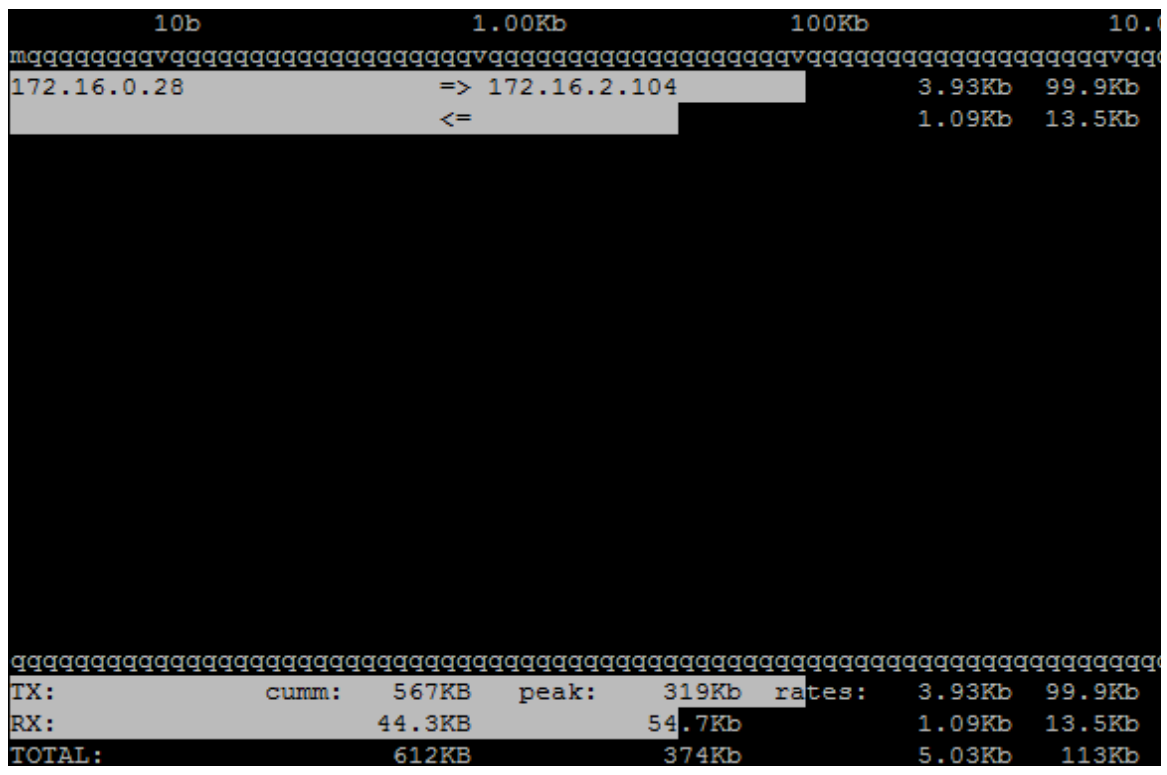

```
        10b                 1.00Kb              100Kb               10.
mqqqqqqqvqqqqqqqqqqqqqqqqqqqvqqqqqqqqqqqqqqqqqqqvqqqqqqqqqqqqqqqqqqqvqq
172.16.0.28                => 172.16.2.104                3.93Kb  99.9Kb
                           <=                             1.09Kb  13.5Kb

qqqqqqqqqqqqqqqqqqqqqqqqqqqqqqqqqqqqqqqqqqqqqqqqqqqqqqqqqqqqqqqqqqqqqqq
TX:             cumm:   567KB   peak:    319Kb  rates:    3.93Kb  99.9Kb
RX:                    44.3KB           54.7Kb            1.09Kb  13.5Kb
TOTAL:                  612KB            374Kb            5.03Kb   113Kb
```

*Рис. 3.* Сетевая активность (один пользователь)

В системе LMS Moodle работает один пользователь. Параметры системы следующие:

- Загруженность процессора 7,3 % от его мощности;
- Использование памяти: 462 Мбайт занято 42 Мбайт свободно;
- Активные процессы: httpd 6,7 % CPU, 3,3 % Mem; mysqld 0,7 % CPU, 3,4 % Mem; java 0,3 % CPU, 21,9 % Mem;

Сетевая активность: скорость передачи 319 Кбит/сек, скорость приема 54.7 Кбит/сек, всего передано за сеанс 612 Кбайт.

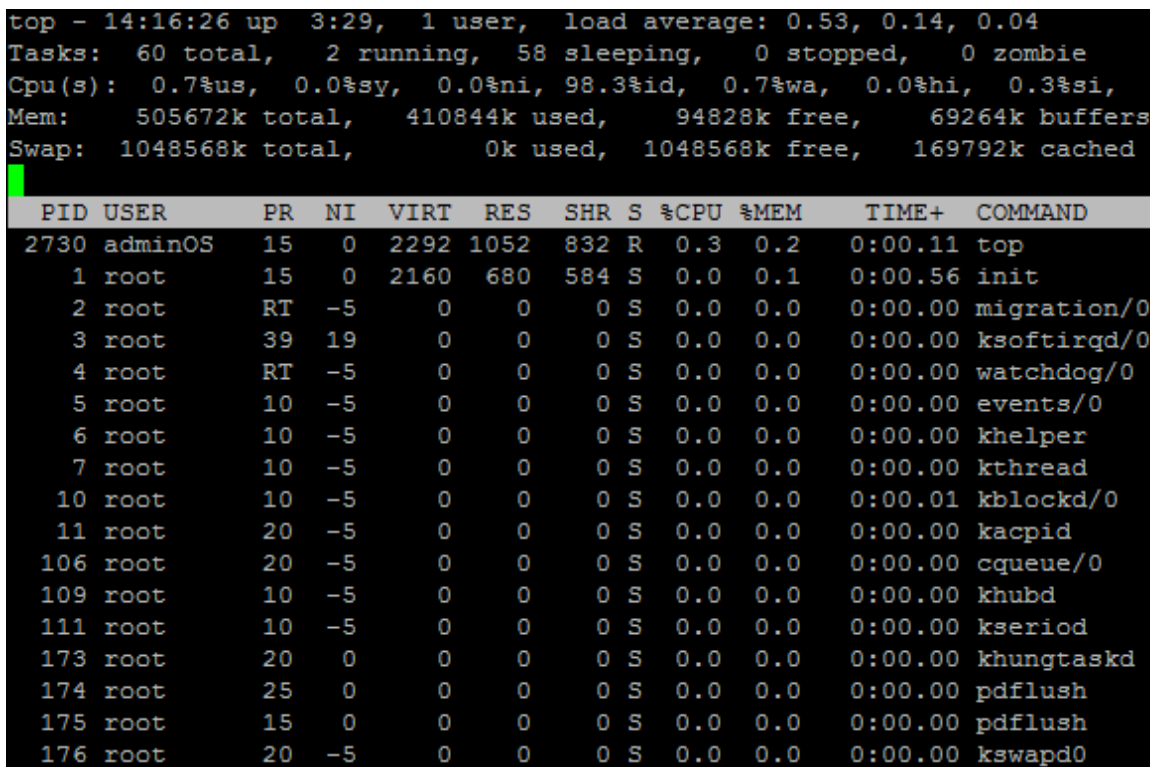

```
top - 14:16:26 up  3:29,  1 user,  load average: 0.53, 0.14, 0.04
Tasks:  60 total,   2 running,  58 sleeping,   0 stopped,   0 zombie
Cpu(s):  0.7%us,  0.0%sy,  0.0%ni, 98.3%id,  0.7%wa,  0.0%hi,  0.3%si,
Mem:    505672k total,   410844k used,    94828k free,    69264k buffers
Swap:  1048568k total,        0k used,  1048568k free,   169792k cached

  PID USER      PR  NI  VIRT  RES  SHR S %CPU %MEM    TIME+  COMMAND
 2730 adminOS   15   0  2292 1052  832 R  0.3  0.2   0:00.11 top
    1 root      15   0  2160  680  584 S  0.0  0.1   0:00.56 init
    2 root      RT  -5     0    0    0 S  0.0  0.0   0:00.00 migration/0
    3 root      39  19     0    0    0 S  0.0  0.0   0:00.00 ksoftirqd/0
    4 root      RT  -5     0    0    0 S  0.0  0.0   0:00.00 watchdog/0
    5 root      10  -5     0    0    0 S  0.0  0.0   0:00.00 events/0
    6 root      10  -5     0    0    0 S  0.0  0.0   0:00.00 khelper
    7 root      10  -5     0    0    0 S  0.0  0.0   0:00.00 kthread
   10 root      10  -5     0    0    0 S  0.0  0.0   0:00.01 kblockd/0
   11 root      20  -5     0    0    0 S  0.0  0.0   0:00.00 kacpid
  106 root      20  -5     0    0    0 S  0.0  0.0   0:00.00 cqueue/0
  109 root      10  -5     0    0    0 S  0.0  0.0   0:00.00 khubd
  111 root      10  -5     0    0    0 S  0.0  0.0   0:00.00 kseriod
  173 root      20   0     0    0    0 S  0.0  0.0   0:00.00 khungtaskd
  174 root      25   0     0    0    0 S  0.0  0.0   0:00.00 pdflush
  175 root      15   0     0    0    0 S  0.0  0.0   0:00.00 pdflush
  176 root      20  -5     0    0    0 S  0.0  0.0   0:00.00 kswapd0
```

*Рис. 4.* Вид системы (работают 5 пользователей)

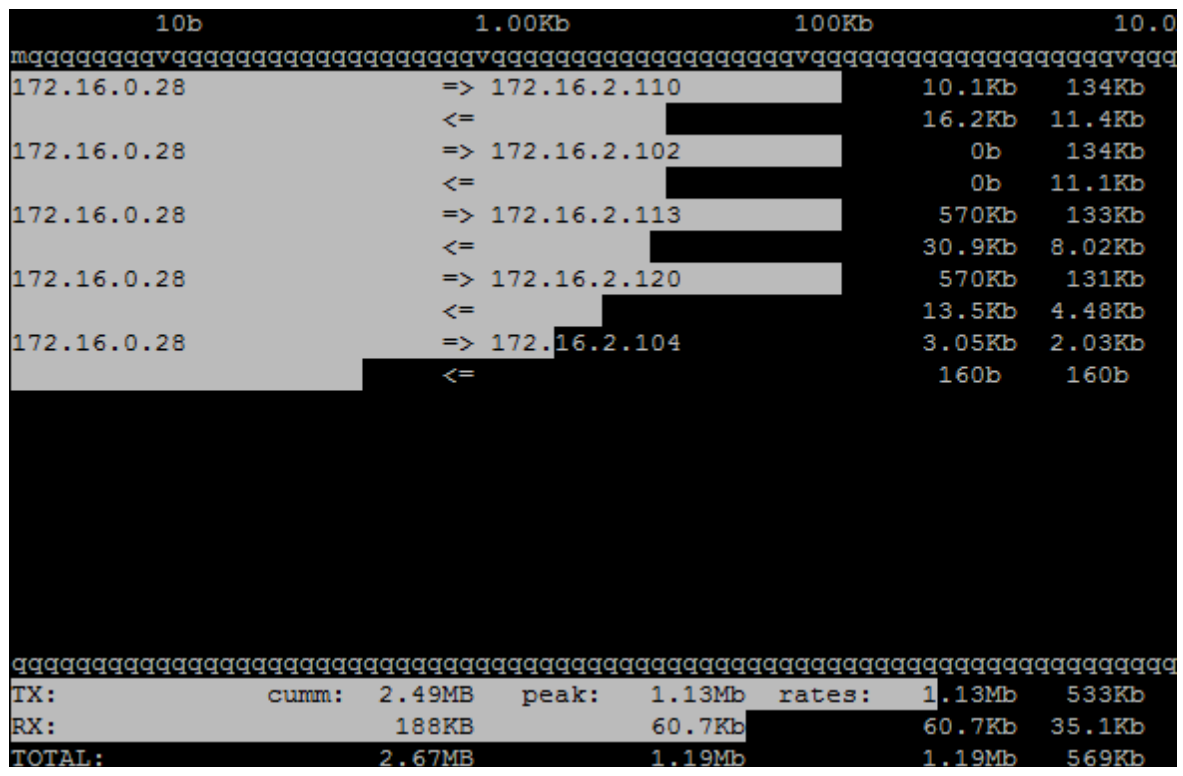

```
        10b                 1.00Kb              100Kb               10.0
mqqqqqqqvqqqqqqqqqqqqqqqqqqqvqqqqqqqqqqqqqqqqqqqvqqqqqqqqqqqqqqqqqqqvqqq
172.16.0.28                => 172.16.2.110                10.1Kb   134Kb
                           <=                             16.2Kb  11.4Kb
172.16.0.28                => 172.16.2.102                    0b   134Kb
                           <=                                 0b  11.1Kb
172.16.0.28                => 172.16.2.113                 570Kb   133Kb
                           <=                             30.9Kb  8.02Kb
172.16.0.28                => 172.16.2.120                 570Kb   131Kb
                           <=                             13.5Kb  4.48Kb
172.16.0.28                => 172.16.2.104                3.05Kb  2.03Kb
                           <=                              160b    160b

qqqqqqqqqqqqqqqqqqqqqqqqqqqqqqqqqqqqqqqqqqqqqqqqqqqqqqqqqqqqqqqqqqqqqqqq
TX:             cumm:  2.49MB   peak:   1.13Mb  rates:    1.13Mb   533Kb
RX:                     188KB           60.7Kb            60.7Kb  35.1Kb
TOTAL:                 2.67MB           1.19Mb            1.19Mb   569Kb
```

*Рис.* 5. Сетевая активность (5 пользователей)

В системе LMS Moodle работают пять пользователей. Параметры системы следующие:

- Загруженность процессора 13 % от его мощности;
- Использование памяти: 410 Мбайт занято, 94 Мбайт свободно;
- Активные процессы: httpd 6,7 % CPU, 10,3 % Mem; mysqld 2,7 % CPU, 3,4 % Mem; java 0,3 % CPU, 21,9 % Mem;
- Сетевая активность: скорость передачи 1,13 Мбит/сек, скорость приема 60,7 Кбит/сек, всего передано за сеанс 2,67 Мбайт.

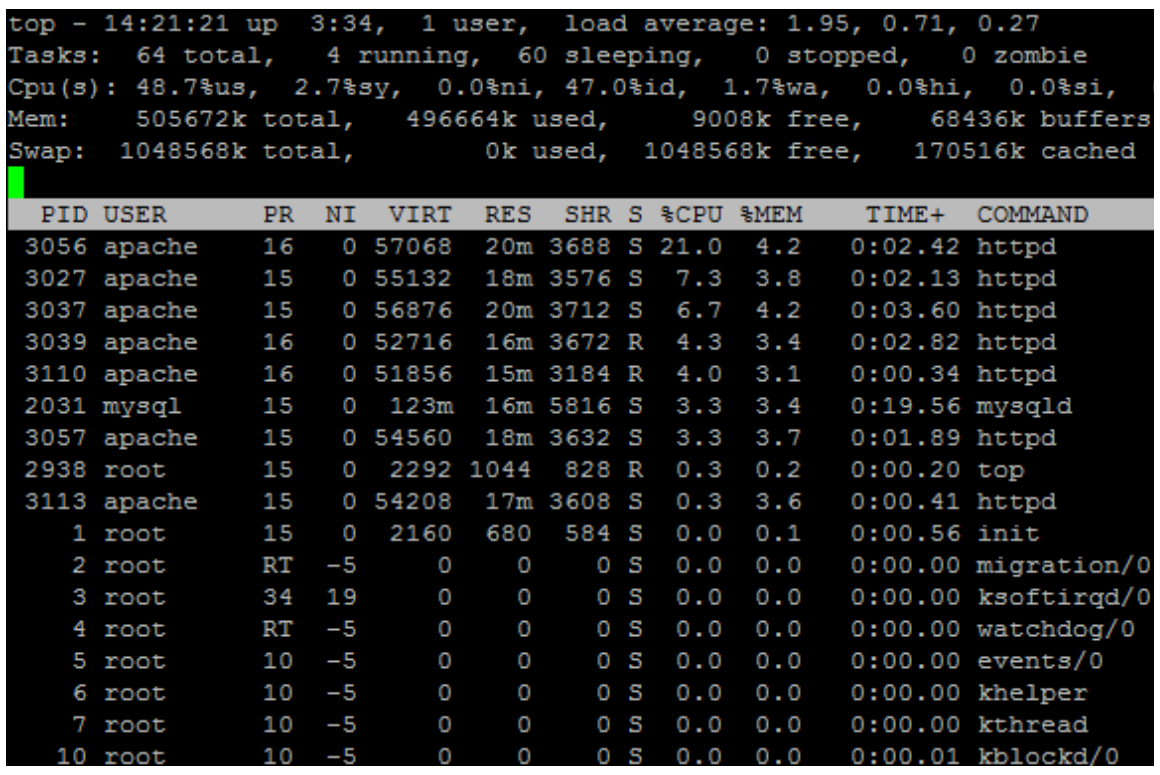

*Рис. 6*. Вид системы (работают 10 пользователей)

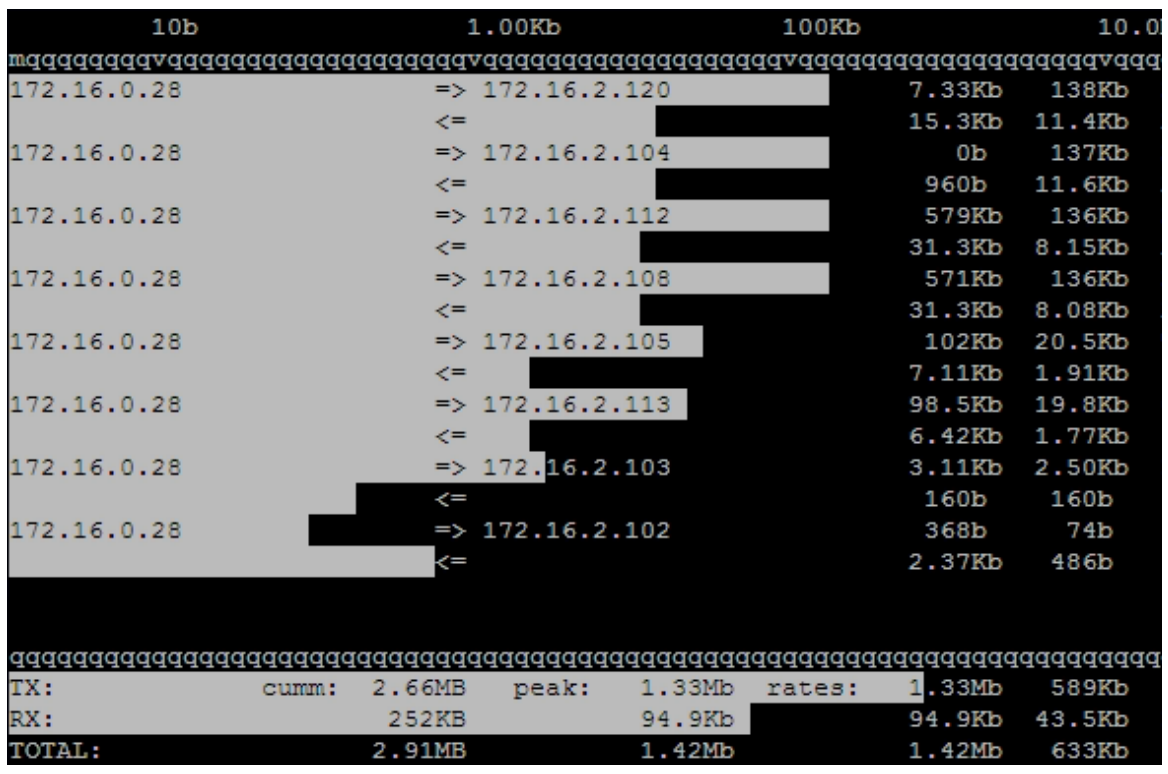

*Рис. 7*. Сетевая активность (работают 10 пользователей)

В системе LMS Moodle работают 10 пользователей. Параметры системы следующие:

- Загруженность процессора 48 % от его мощности;
- Использование памяти: 496 Мбайт занято, 9 Мбайт свободно;
- Активные процессы: httpd 45 % CPU, 18 % Mem; mysqld 3,3 % CPU, 3,4 % Mem; java 0 % CPU, 21,9 % Mem;
- Сетевая активность: скорость передачи 1,33 Мбит/сек, скорость приема 94,6 Кбит/сек, всего передано за сеанс 2,91 Мбайт.

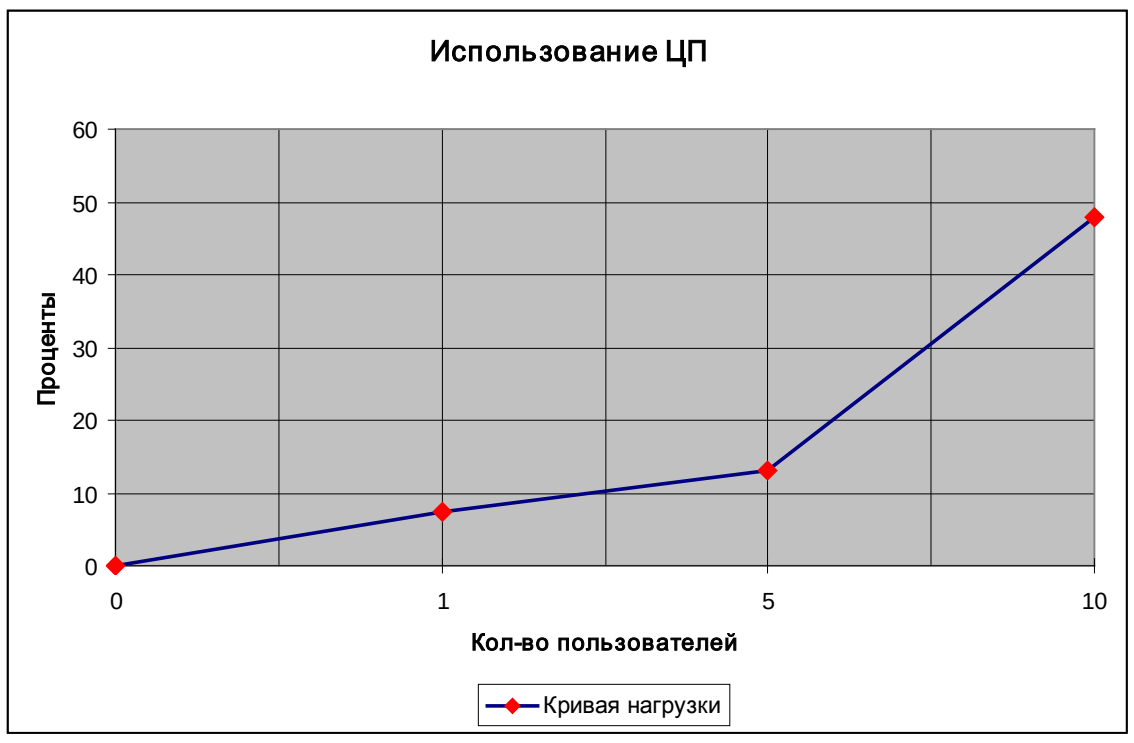

*Рис. 8*. Загруженность процессора

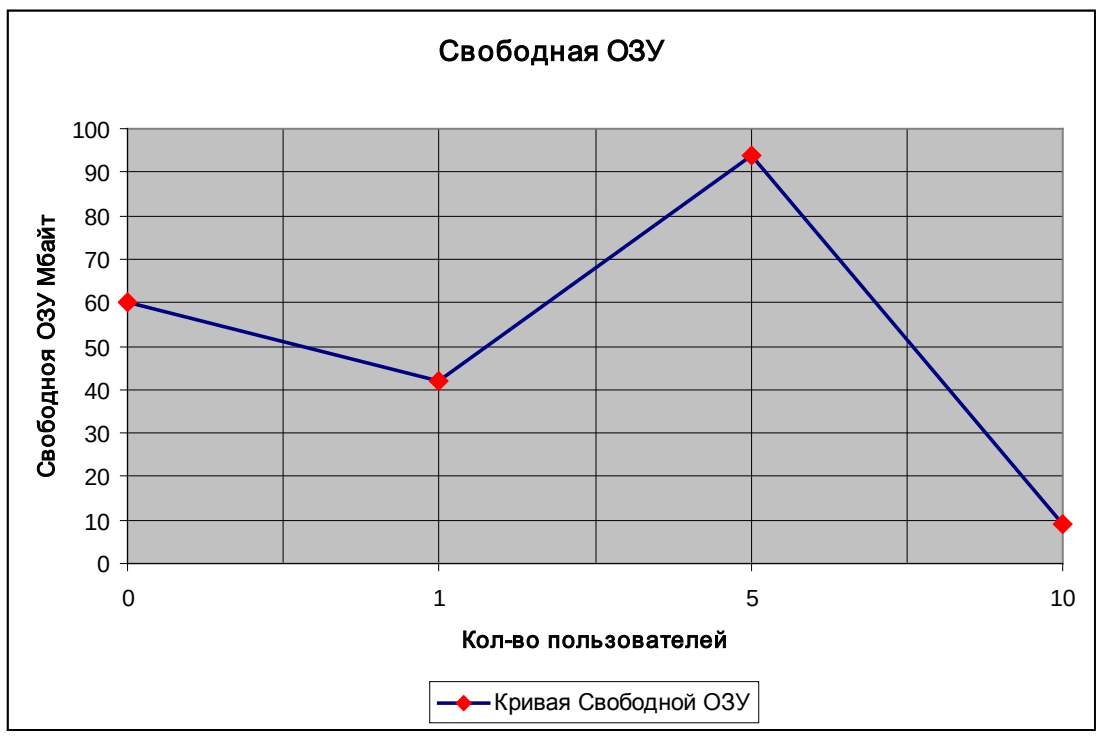

*Рис. 9*. Свободная ОЗУ

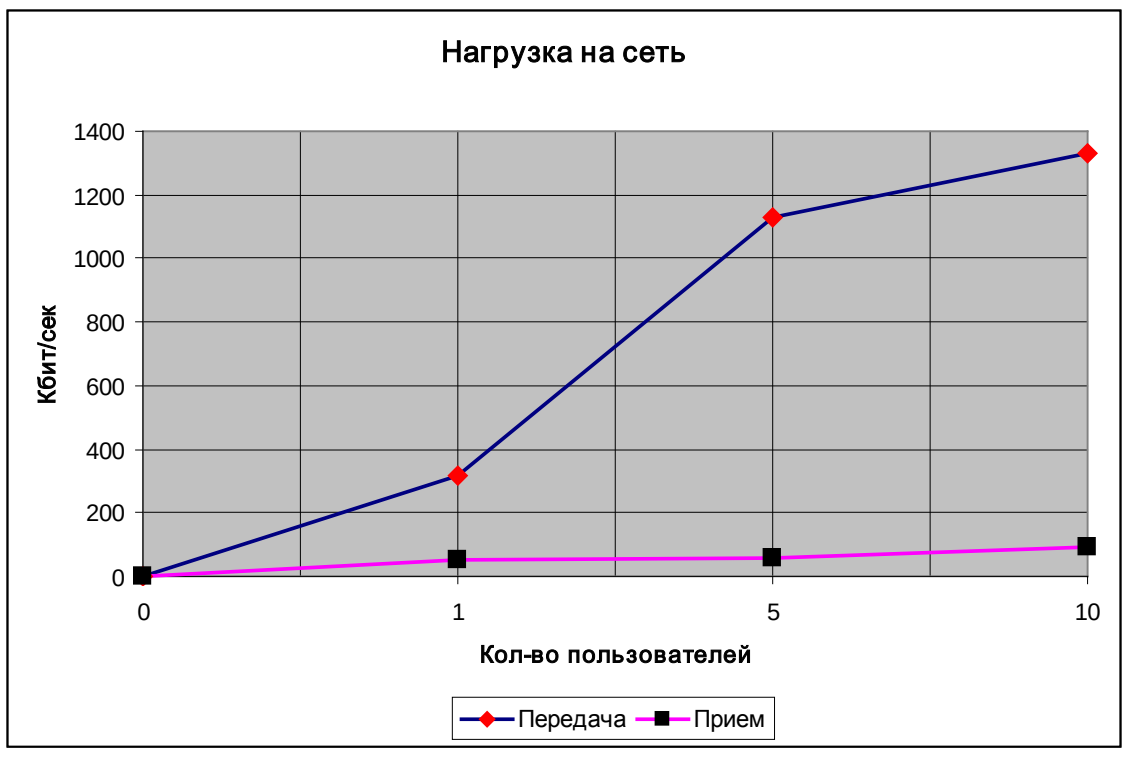

*Рис. 10*. Нагрузка на сеть

На базе беспроводной сети конференцию удалось без проблем реализовать с помощью Openmeetings с 5 пользователями. Тестирование на базе проводной сети 100 Мбит/сек, проводилось с десятью пользователями. Результаты представлены на рис. 11-12.

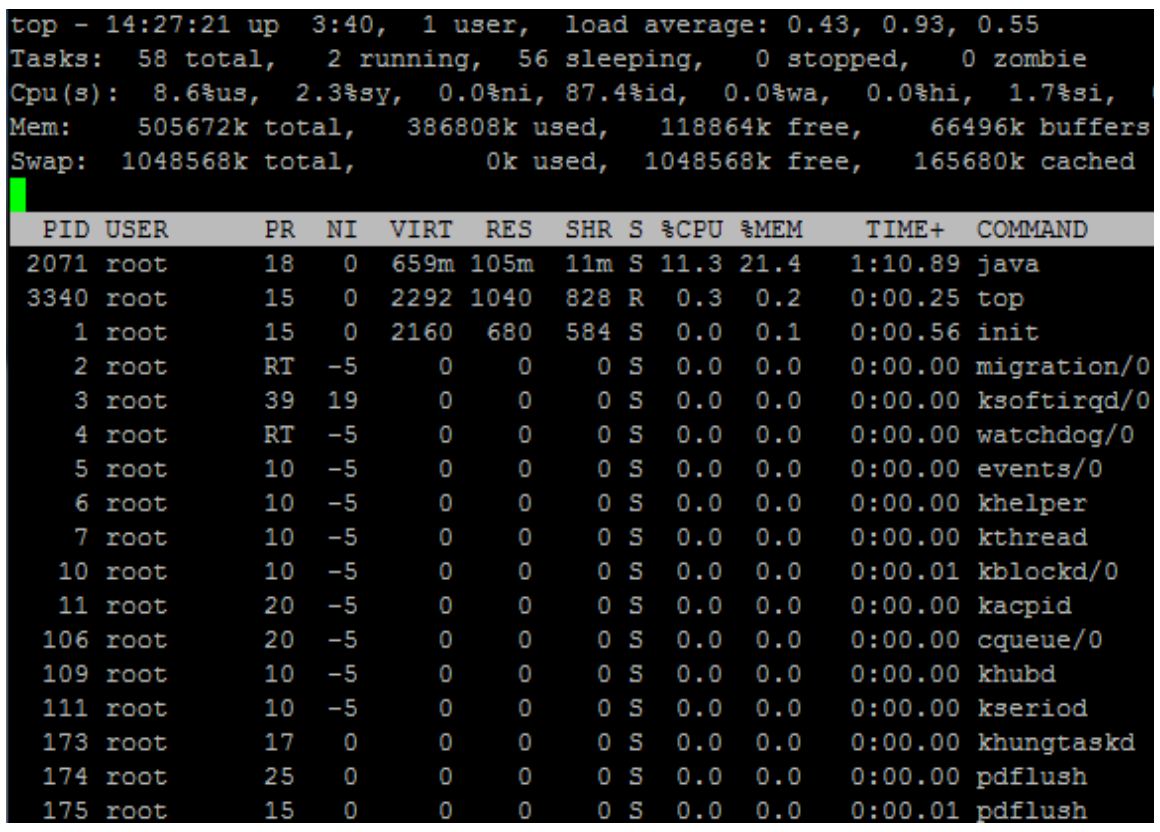

*Рис. 11.* Вид системы
(работают 5 пользователей)

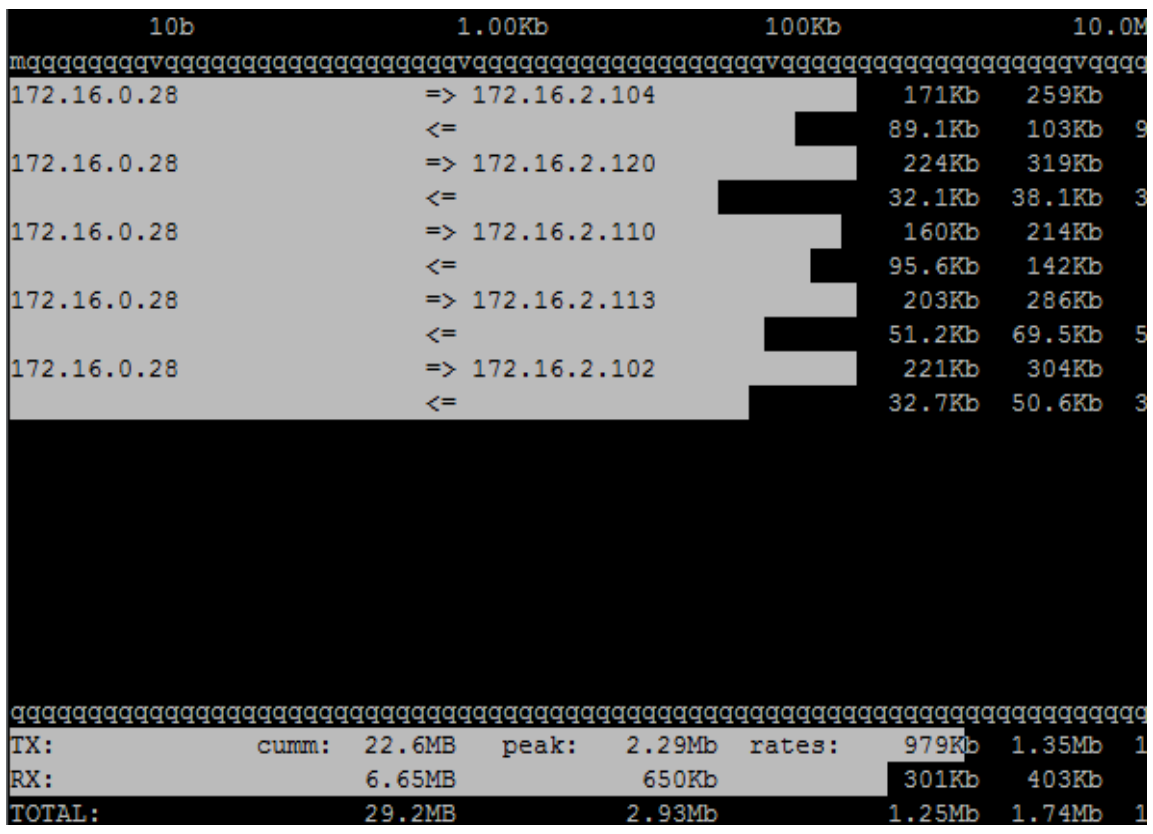

*Рис. 12.* Сетевая активность
(работают 5 пользователей)

В системе Openmeetings работают 5 пользователей. Параметры системы следующие:

- Загруженность процессора 8,6 % от его мощности;
- Использование памяти: 386 Мбайт занято 118 Мбайт свободно;
- Активные процессы: java 11,6 % CPU, 21,4 % Mem;
- Сетевая активность: скорость передачи 2,29 Мбит/сек, скорость приема 650 Кбит/сек, всего передано за сеанс 29,2 Мбайт.

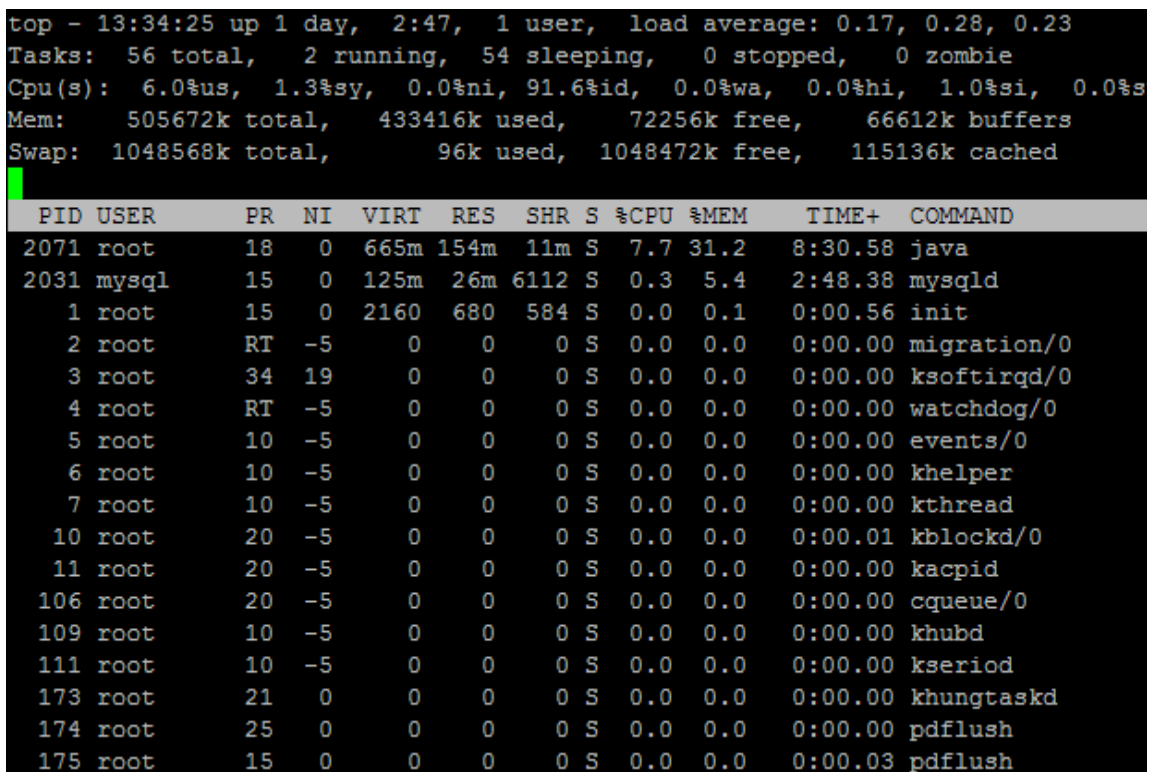

*Рис. 11.* Вид системы
(работают 10 пользователей)

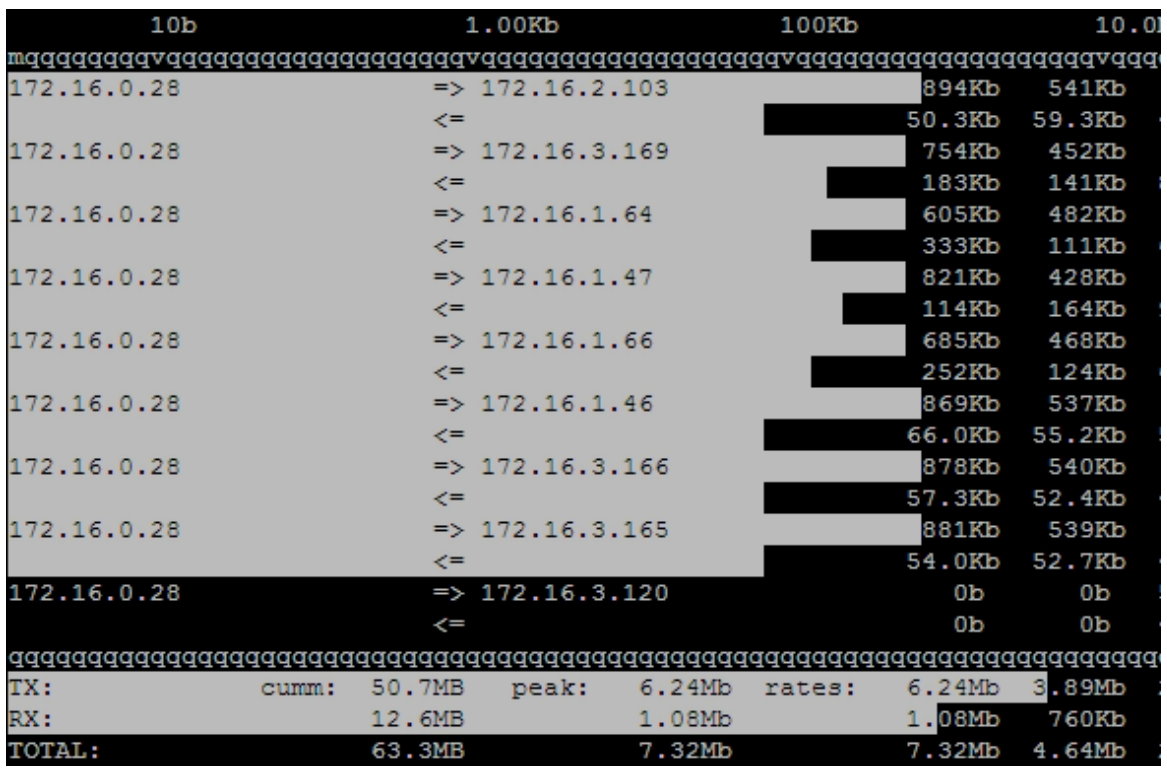

*Рис. 12.* Сетевая активность
(работают 10 пользователей).

В системе Openmeetings работают 10 пользователей. Параметры системы следующие:

- Загруженность процессора 6 % от его мощности;
- Использование памяти: 433 Мбайт занято 72 Мбайт свободно;
- Активные процессы: java 7,7 % CPU, 31,2 % Mem; mysqld 0,3 % CPU, 5,4 % Mem;
- Сетевая активность: скорость передачи 6,24 Мбит/сек, скорость приема 1,08 Мбит/сек, всего передано за сеанс 63,3 Мбайт.

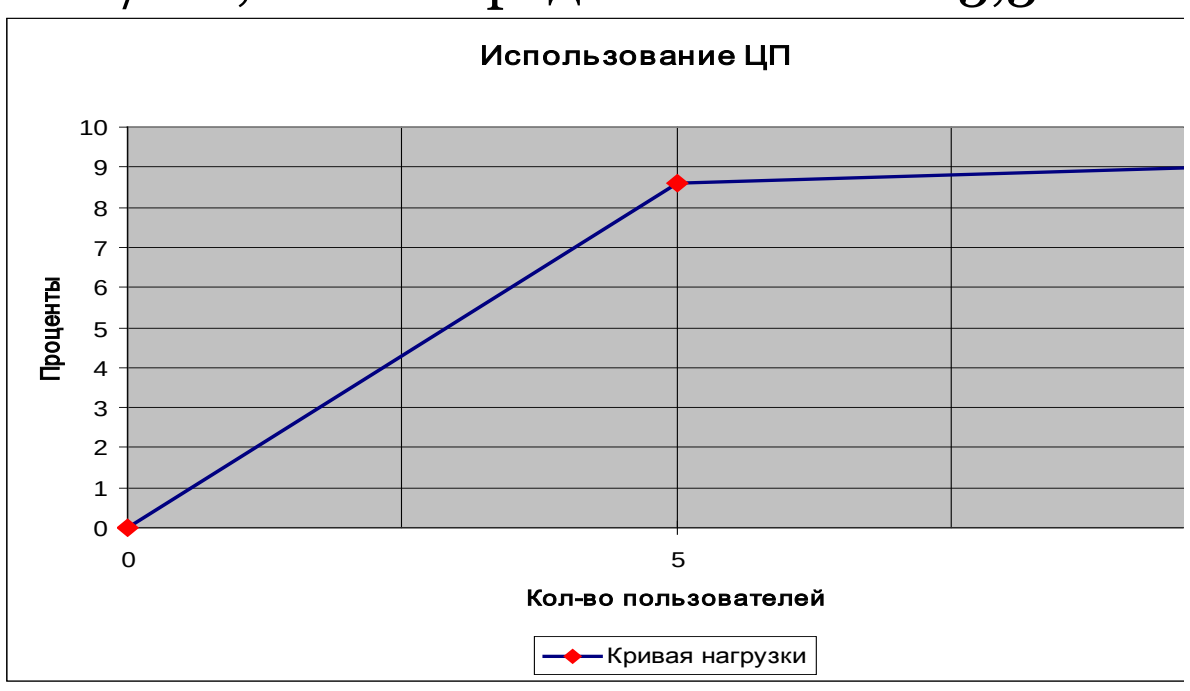

*Рис. 15.* Нагрузка на процессор

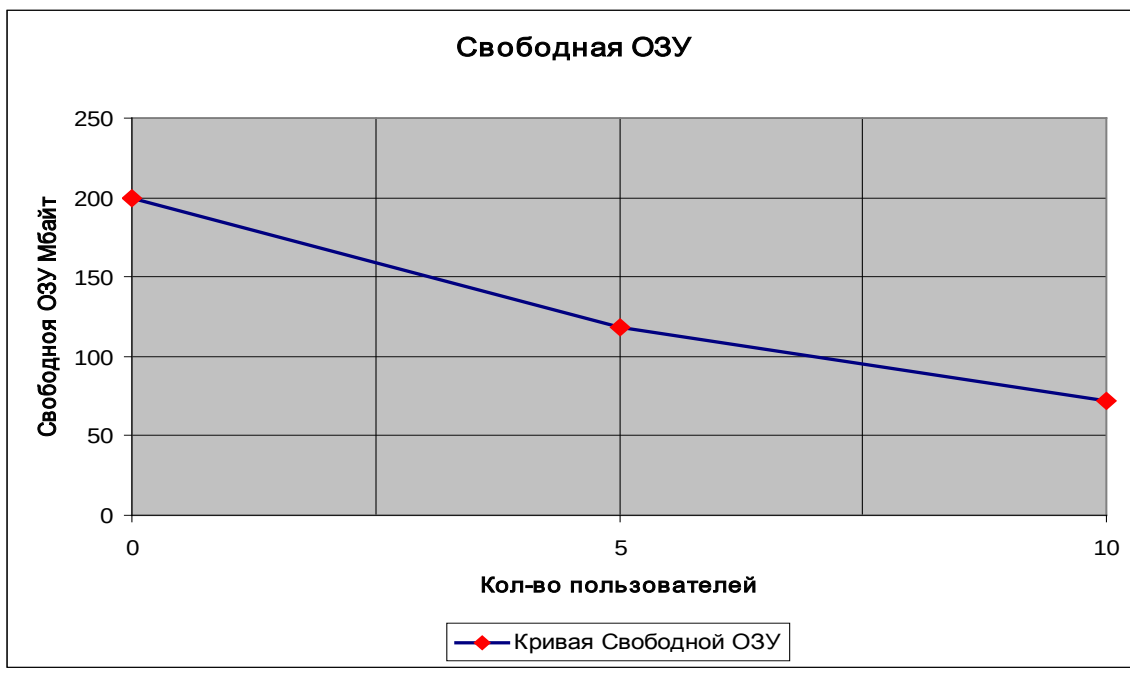

*Рис. 16.* Свободная память системы

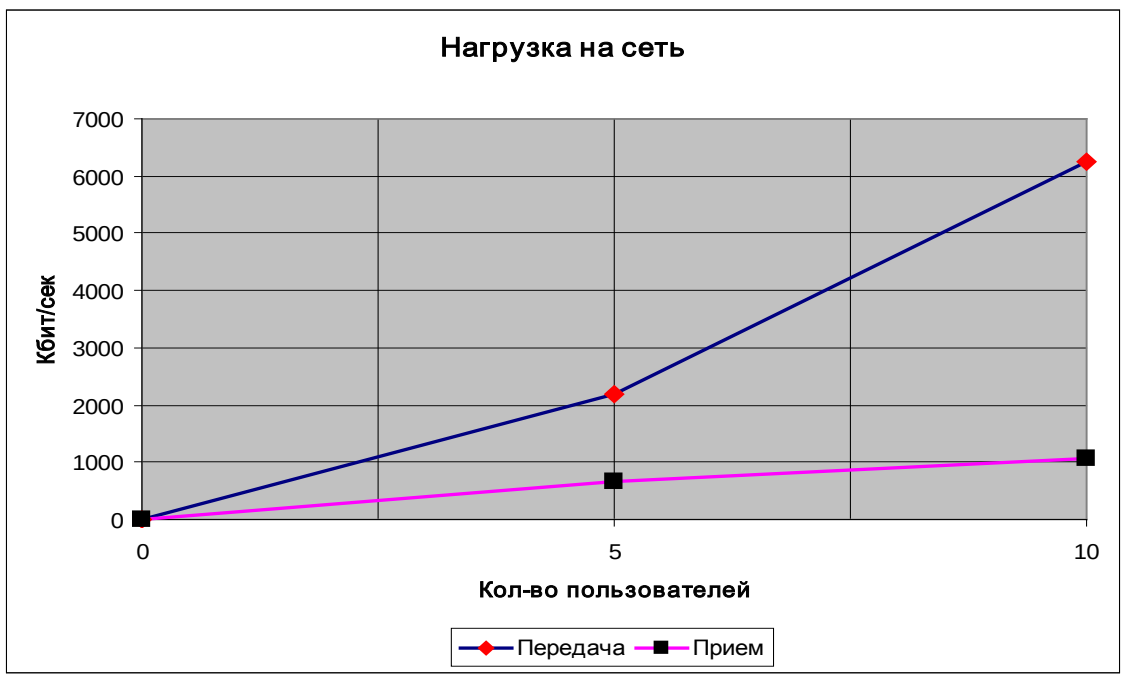

*Рис. 17.* Нагрузка на сеть

**Заключение.** Таким образом, система LMS Moodle работает стабильно и без сбоев и даже с учетом небольшой свободной оперативной памяти система вполне выдержит нагрузку в 30-50 пользователей. Для большего количества клиентов требуется увеличение свободной оперативной памяти (как минимум до 1 Гбайт). Во время тестирования нагрузка на процессор изменялась в зависимости от выбора параметров технического сопровождения. Пик нагрузки приходился на максимальное число запросов от клиентов.

Система организации веб-взаимодействия в режиме конференции и с использованием Wi-Fi сети позволила организовать стабильную работу 5 пользователей, после подключения 6-го пользователя начинались зависания компьютеров-клиентов, и после подключения 8-го пользователя происходила полная неработоспособность конференции. Анализ загруженности сервера и степень свободных ресурсов позволили сделать вывод о том, что в данном случае оказалось недостаточно ресурсов среды передачи данных. Для оптимальной работы системы требуется канал со следующими характеристиками:

Для сервера:

- Входящий: 256 кбит/сек * N;
- Исходящий: 256 кбит/сек * (N – 1).

Для клиентов:

- Входящий: 256 кбит/сек * (N – 1);
- Исходящий: 256 кбит/сек,

где N – количество пользователей.

Также нужно менее 80 мс времени для прохождения пакетов (ping) между конечными узлами и от каждой из точек до сервера.

Сеть Wi-Fi при подключении 8 компьютеров к серверу создавала трафик, равный 17 Мбит/сек, точка доступа не справлялась с такой нагрузкой, поэтому появлялись большие задержки.

При втором тестировании использовалась проводная сеть Ethernet с пропускной способностью, равной 100 Мбит/сек. Подключались 10 клиентов, присутствовала задержка менее 1 секунды, нагрузка на сервер оставалась менее 10 % его мощности. К данному серверу могут быть подключены одновременно 15-20 пользователей, однако, их большее количество вызовет переполнение оперативной памяти и работа замедлится.

В результате экспериментальной работы проведено тестирование созданного сервера. В качестве клиентов использовались ноутбуки HP Compaq 610. Система Openmeetings базируется на сервере Red5, который управляет работой потоков видео и аудио. При этом видео кодируется на компьютере-клиенте, что создает некоторую нагрузку на процессор. В упомянутой модели ноутбука используется процессор AMD Turion II Dual-Core 2400 МГц, нагрузка на который составила 25 %, на ноутбуке установлена операционная система Microsoft Windows 7 Home Basic. Эта нагрузка не зависит от количества подключенных пользователей.

Во время всего тестирования нагрузка на сервер не превышала 50 % от его мощности, система не использовала дополнительную память из раздела SWAP.

Для комфортной работы LMS Moodle 512 Мбайт оперативной памяти вполне достаточно. Система выдержит большое (до 100) количество пользователей. Путем изменения настроек сервера (отключены некоторые службы веб-сервера и компоненты системы) можно добиться подключения большого числа пользователей.

Для работы LMS Moodle и Openmeetings на одном сервере оперативной памяти 512 Мбайт явно недостаточно. Требуется увеличение объема ОЗУ хотя бы до 1024 Мбайт. К тому же при отправке файла на доску (совместный доступ к экрану или отдельным приложениям), участники конференции испытывали трудности, так как происходит конвертация документа в формат swf. Выходом из этой ситуации служит использование многоядерного процессора, который способен обработать конвертацию документа и не допустить задержек для пользователей.

Сервер видеоконференций должен обладать каналом, обеспечивающим достаточную скорость и малое время задержки от каждого клиента до сервера.

**Примечания:**

## Server Component Installation and Testing of the University Information and Educational Environment on the Moodle LMS Platform

[1] Irina Eremina
[2] Aleksandr Rozentsvaig
[3] Rushan Ziatdinov

[1-2] Naberezhnye Chelny Institute of Kazan (Volga Region) Federal University,
Naberezhnye Chelny, Republic of Tatarstan, Russian Federation
[1] PhD, Associate Professor
E-mail: ereminaii@yandex.ru
[2] Doctor of Technical Sciences, Professor
E-mail: a_k_r@mail.ru
[3] Fatih University, Istanbul, Turkey
PhD, Assistant Professor
E-mail: ziatdinov@fatih.edu.tr, rushanziatdinov@yandex.ru

**Abstract.** The informational educational environment (IEE) of an institution is a complex multilevel system which, along with methodical, organizational and cultural resources, accumulates the intellectual and technical potential of a university, as well as the informative and activity components of the learners and teachers. In practice, the formation of IEE is actually based on the creation of information technologies and their integration into the existing educational environment of the institution. The management of this system is carried out using specialized equipment and software. For the successful formation and operation of IEE, in the present work we review software products that form the basis of the organization of interactive and web interactions between students, teachers and all participants of the educational process. We analyse the technical capabilities that have provided users with IEE services such as the Apache web server with connected modules PHP, MySQL, the Java virtual machine and the Red5 server. We demonstrate the possibility of obtaining results from the interaction of these products, and reports on users' work in webinars, video conferences and web conferences.